\documentclass[10pt,twocolumn]{article}
\usepackage{times} 
\usepackage{xspace}

\usepackage{amsmath}
\usepackage{graphicx}
\usepackage{booktabs} 

\usepackage{url,enumitem}
\usepackage{xcolor}

\newcommand{\reffigure}[1]{Figure~\ref{#1}}
\newcommand{\btree}{B$^+$-tree\xspace}

\date{}

\begin{document}
\title{Efficiently Reclaiming Space in a Log Structured Store}
\author{
David Lomet \\
Microsoft Research \\
Redmond, WA 98052 \\
\texttt{lomet@microsoft.com}
\and 
Chen Luo \\
University of California, Irvine \\
Irvine, CA 92697 \\
\texttt{cluo8@uci.edu}
}

\maketitle

\begin{abstract}
A log structured store uses a single write I/O for a number of diverse and non-contiguous pages within a large buffer instead of using a write I/O for each page separately. This requires that pages be relocated on every write, because pages are never updated in place. Instead, pages are dynamically remapped on every write. Log structuring was invented for and used initially in file systems. Today, a form of log structuring is used in SSD controllers because an SSD requires the erasure of a large block of pages before flash storage can be reused. No update-in-place requires that the storage for out-of-date pages be reclaimed (garbage collected or ``cleaned''). We analyze cleaning performance and introduce a cleaning strategy that uses a new way to prioritize the order in which stale pages are garbage collected. Our cleaning strategy approximates an ``optimal cleaning strategy''. Simulation studies confirm the results of the analysis. This strategy is a significant improvement over previous cleaning strategies.
\end{abstract}

\section{Introduction}

\subsection{LFS and FTL Similarities}

The cost of executing I/O and the maximum number of I/Os that are provided in a system is important for both system cost and performance. Log structured file systems (LFS)~\cite{RO92,RO92a} were introduced both to reduce the cost of I/O and to increase the number of blocks that could be written by exploiting batching. An LFS writes batches of diverse pages to secondary storage using large buffers. Further, because the buffer must be written to contiguous secondary storage, LFS's unit of reclaimed storage is a buffer size area on secondary storage. (In a RAID setting~\cite{PGK88}, the large buffer is striped across an array of disks.)

An SSD controller, in implementing its flash translation layer (FTL)~\cite{wikiFTL}, needs also to reclaim large storage units. It writes pages to SSD storage within an erase block, which is the unit of erasure for flash storage. Pages must be erased between writes, so the unit of reclaimed storage is the erase block. How efficiently it does this helps to determine how many IOPS and cost/IOPS that it can provide. 

\begin{figure}
\includegraphics[width=\linewidth]{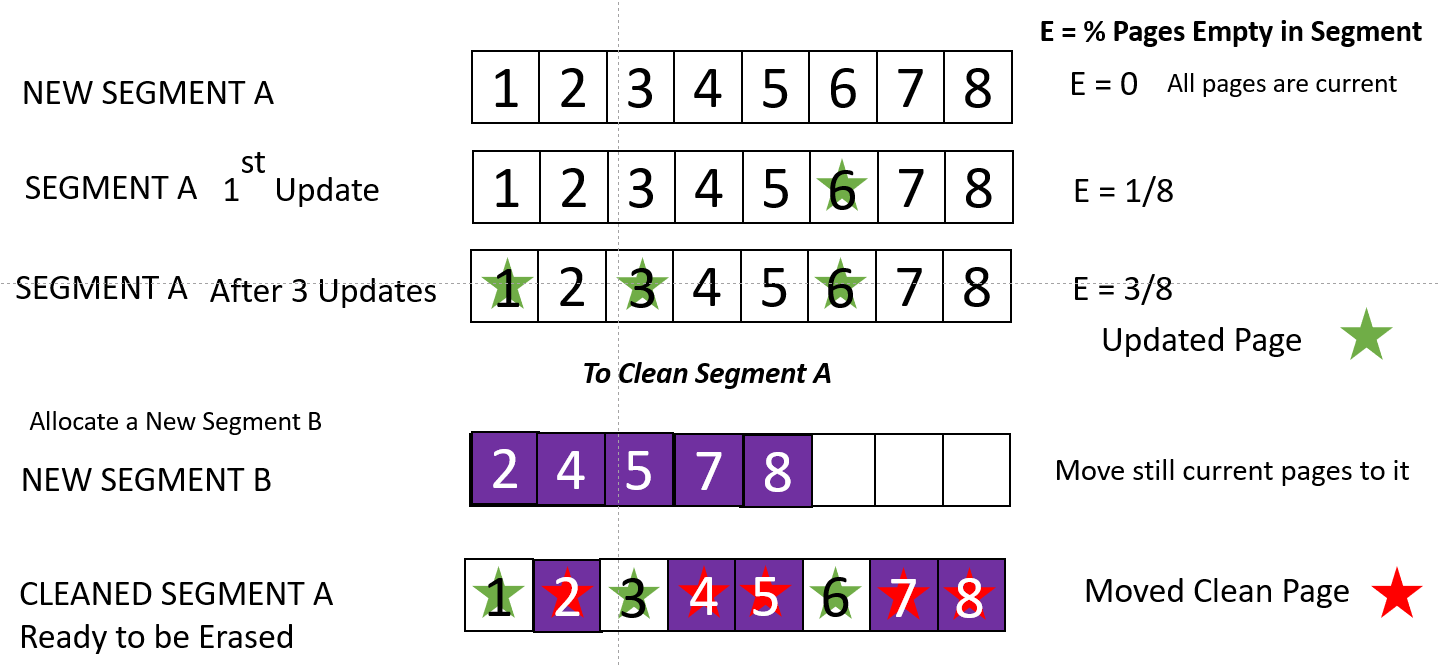}
\caption{Checkboard segment containing current and obsolete pages, and the cleaning process.}
\label{fig:checker}
\end{figure}

So for both LFS and FTL, the unit of reclaimed storage (which we refer to subsequently as a ``segment'') must be an area that contains a number of pages. However, it is the pages themselves that are the unit of obsolescence. That is, an update to a page replaces the page's prior version, while the rest of its containing segment is unchanged by the update. Several updates can, of course, make other pages in the segment obsolete as well. This leads to segments that have a ``checkerboard" pattern of over-written (and hence obsolete) pages and pages that continue to represent the current content of their respective pages. Figure~\ref{fig:checker} illustrates this. 

\subsection{Cleaning}

Whether it is an LFS or an SSD's FTL, neither do update-in-place, and both require that out-of-date versions of pages be garbage collected (called ``cleaning'' in \cite{RO92}). Without cleaning, secondary storage fills up with partially overwritten segments, i.e. segments in which some of the pages are current versions and some are obsolete ({\bf empty} of current versions) and hence garbage. And importantly, new storage for writes needs to be acquired in segment size units to ensure the system's ability to continue writing new versions of pages to storage.

Obviously, it would be great if we could find a segment in which all pages have been over-written and are hence empty. Then we could simply select this segment for re-use, and erase it if it is a flash erase block. But this is incredibly unlikely as we are usually dealing with segments capable of storing a few hundred pages or more. So we have to deal with segments that have a checkerboard of current and empty pages. The strategy to do this entails writing the still current pages to a new segment. Once these pages are written elsewhere (somewhat as if they were being updated by users), all pages in the segment are then empty. This cleaning process has emptied the pages that were current when cleaning began for the segment. Figure~\ref{fig:checker} shows how segments become increasingly dirty, and how cleaning moves the still current pages to another segment so that there are no remaining current versions of pages in the original segment.

How efficiently we can perform cleaning is an important factor in how well these systems perform. This cost is measured in the number of page moves required to perform cleaning, which is the number of pages that still contain current data. There are two reasons why this ``efficiency'' matters.
\begin{enumerate}[leftmargin=.2in,itemsep=-3pt]
\item The number of page moves is directly related to the execution cost of garbage collection. Execution cycles spent on cleaning are cycles that are unavailable for tasks more directly related to user I/O operations. 
\item For SSDs, the number of page moves is related to what is called ``write amplification'' \cite{wikiWA}. This is a measure of how many writes result from a single page user write. The larger the write amplification, the more wear on the flash memory, which is limited in the number of writes that it can provide.
\end{enumerate}

\subsection{This Paper and our Contributions}

The remainder of the paper is organized as follows.
\begin{itemize}[leftmargin=.2in,itemsep=-3pt]
\item Section~\ref{sec:clean} starts by providing an algebraic derivation of cleaning cost, originally derived in~\cite{RO92}. Following that is a derivation of simple round robin, age-based cleaning costs, previously derived in~\cite{Lo94,AM10,DBB15,D12}. This age based cleaning analysis is used later.
\item Section~\ref{sec:skewed} contains one of the significant contributions of this work. We present an analysis demonstrating that a skewed data and update distribution with hot and cold data of different sizes can be exploited to reduce cleaning cost. While this is known~\cite{RO92}, our new analysis convincingly demonstrates this and provides a lower bound on cleaning cost for various skews. 
\item How to order segments for cleaning so as to produce the lowest cost is the subject of Section~\ref{sec:order}. Here we introduce the major contribution of the paper, the ``optimal'' approach to choosing the order of cleaning among the in-use segments. This analysis produces a formula that can be evaluated to determine the cleaning order for segments that are in-use.
\item The analysis of the prior section needs to be converted into an executable cleaning program. This is described in Section~\ref{sec:how} and is another contribution of the paper.
\item Our cleaning simulation is described in Section~\ref{sec:eval}. This, our last contribution, demonstrates that our cleaning order produces the lowest cost cleaning of the major competing strategies. 
\item The last two sections contain related work and a short discussion.
\end{itemize}
 
\section{Cleaning Segments}
\label{sec:clean}
 
In order to reuse empty pages of a segment, the segment is cleaned. The cleaning process consists of reading the segment and re-writing, hence relocating, its still current (non-empty) pages. The segment can then be re-used in its entirety for writing new versions of pages. The efficiency of this process depends on the number of {\bf empty} pages, i.e., pages containing out-of-date versions, a segment being cleaned contains. The more empty pages, the lower the cleaning cost per page, since these empty frames represent the gain achieved by the cleaning process.
 
The number of empty pages in a segment increases with time. Whenever a page is updated, it is re-written to a new location. The storage (sometimes called the page frame) that contained the page's prior version becomes empty. Only cleaning, which empties all pages of a segment, results in empty pages being usable for future writes. Otherwise, segments become increasingly empty with time.
Note that without cleaning, there is no guarantee that some segments will ever consist of entirely empty pages.

\subsection{The Cost of Cleaning}
 
We want to know how many I/O accesses, including those for cleaning, are required to write a segment of new pages to the 
disk. This will be a function of how empty segments are when cleaned, as once used segments fill the allotted storage, space for new data must be reclaimed from previously written segments. 
 
Let $E$ denote the fraction of a segment that is empty. The reading of a segment for cleaning produces $E*S$ empty pages, where $S$ is the number of pages per segment. In order to write $S$ page frames of new data, $1/E$ segments must be cleaned, i.e. read. Further, for each segment read, $(1 - E)*S$ pages must be written to move these still alive pages to a different segment. Finally, the cleaned segment, now completely available, must be written with the $S$ new pages. Thus, the total I/O cost of writing that segment, which we denote as $Cost_{seg}$ is
\begin{equation}
Cost_{seg} = \frac{1}{E} reads + \frac{1}{E} (1 - E) writes + 1 = \frac{2}{E} 
\label{eq:IOseg}
\end{equation}

Also of importance is the write amplification $Wamp$, which is the number of buffer writes performed compared with the number of buffer writes that the user herself performs. In the $Cost_{seg}$ formula in \ref{eq:IOseg}, this is the second term on the right. Thus,
\begin{equation}
Wamp = \frac{1}{E} (1 - E) 
\label{eq:Wamp}
\end{equation}
We report our simulation results in Section~\ref{sec:eval} where our graphs are in terms of $Wamp$.

$E$ is affected by the fraction of the disk used to store current versions of user data. We call this fraction the fill factor, $F$. The apparent size of the disk for users is $F$ times the real physical size of the disk. Cleaning efficiency depends strongly on $F$. The lower $F$ is, the higher the amount of empty space we are likely to find when cleaning segments. In particular, we can find segments for which $E \ge (1-F)$, as $(1-F)$ is the average amount of empty space over the entire disk when users have completely filled their storage allocation. We should, on average, be able to find larger values of $E$ in segments we clean by careful choice of segments. For example, with a fill factor $F = .8$, on average $E \ge .2$, leading to $IO/seg \le 10$. Hence, in this case, once segment size exceeds $10$ pages, an LFS performs fewer writes per page than a conventional block-at-a-time system.

We want to order the cleaning of segments so as to minimize $Cost_{seg}$. Given equation~\ref{eq:IOseg}, it is clear that this cost is minimized when we maximize the average value for $E$ {\bf at the time a segment is cleaned}. Section~\ref{sec:order} concerns how we prioritize (order) the segments for cleaning, and the impact of that priority ordering on the average $E$ that we see when cleaning segments.
 
\subsection{Age Based Cleaning}

A particularly simple cleaning priority is to always clean the oldest segment, i.e. the one that was written the longest time in the past. This means that we first clean the segment with the largest $age$. It is more convenient for subsequent discussion to think of this as cleaning first the segment with the smallest value of $1/age$. Age-based cleaning leads to a very simple implementation where segments can be organized in a circular buffer. The segment written at the earliest time is the one to be cleaned next. 

A uniform distribution analysis using age based cleaning priority gives a worst case bound on cleaning efficiency which is more accurate than simply using the fill factor $F$. A similar analysis was done earlier~\cite{DBB15}. The disk is assumed to be full, i.e. up to its fill factor $F$. Each page is assumed equally likely to be updated, i.e. have the same probability of update. That is, the update frequency for all pages is $U_{pf} = 1$.  (Note that regardless of update distribution, where $U_{pf}$ may be greater for some pages than for others, average $(U_{pf}) = 1$ over the entire disk is normalized to $1$.) 
 
Let $P$ be the number of pages of the user visible disk. Let N be the number of page writes between the time that a segment is written and when it is cleaned. Then $E$ is the probability that a page written subsequently ``collides" with a page in the segment. That is
\begin{equation}
E = 1 - (\frac{P - 1}{P})^N 
\end{equation}

We are given P. The difficulty is in determining the value for $N$ since $N$ depends upon write cost, i.e. how efficient cleaning is. The physical disk contains $P/F$ pages, and hence $(P/F)/S$ physical segments. For age based priority, we will have written all the physical segments (less one) before cleaning a segment. Each of those segments will have $E*S$ pages available for writing new data. Hence,
\[
N = \frac{(P/F)}{S} E S = \frac{P E}{F}
\]
Then, we have
\[
E = 1 - (\frac{P - 1}{P})^{P\frac{E}{F}}
\]
 
The above equation is recursive, but a fixpoint can be found numerically. Once $P$ is sufficiently large, e.g. greater than 30, this result depends almost entirely on the value of $F$. In fact, in the limit, as $P$ goes to infinity,
\begin{equation}
E = 1 - (\frac{1}{e})^{\frac{E}{F}}
\label{eq:E}
\end{equation}
because
\[
\lim_{P \rightarrow \infty} (\frac{P - 1}{P})^{P} = \frac{1}{e}
\]
The results for some values for $F$ are
presented in Table~\ref{tab:clean}. 

\begin{table}
\begin{footnotesize}
\begin{center}
\begin{tabular}{|c|c|c|c|c|c|c|}
\hline
  F  & 1-F & E & MDC-opt & Cost & R=E/(1-F) & Wamp \\
\hline
 .975 & .025 & .048 & .048  & 41.7 & 1.94 & 19.8 \\
 .95 & .05 & .094 & .097 & 21.3 & 1.92 & 9.64 \\
 .90 & .10 & .19 & .192 & 10.5 & 1.92 & 4.26 \\
 .85 & .15 & .29 & .283 & 6.90 & 1.90 & 2.45 \\
 .80 & .20 & .375 & .370 & 5.33 & 1.88 & 1.66 \\
 .75 & .25 & .45 & .453 & 4.44 & 1.80 & 1.22 \\
 .70 & .30 & .53 & .532 & 3.78 & 1.77 & .887 \\
 .65 & .35 & .60 & .606 & 3.33 & 1.71 & .666 \\
 .60 & .40 & .67 & .675 & 2.99 & 1.68 & .493 \\
 .55 & .45 & .74 & .738 & 2.70 & 1.64 & .351 \\
 .50 & .50 & .80 & .796 & 2.50 & 1.60 & .250 \\
 .45 & .55 & .85 & .847 & 2.35 & 1.55 & .176 \\
 .40 & .60 & .89 & .892 & 2.24 & 1.49 & .124 \\
 .35 & .65 & .93 & .929 & 2.15 & 1.43 & .075 \\
 .30 & .70 & .96 & .959 & 2.08 & 1.37 & .042 \\
 .25 & .75 & .98 & .980 & 2.04 & 1.31 & .020 \\
 .20 & .80 & .993 & .993 & 2.014 & 1.24 & .007\\
\hline
\end{tabular}
\end{center}
\caption{Fill Factor F and 1- F vs Segment Emptiness When Cleaned} 
\label{tab:clean}
\end{footnotesize}
\end{table}

This table, all columns except MDC-opt\footnote{MDC-opt is the simulation result for the minimum declining cost (MDC) algorithm proposed later in the paper. This column can be ignored for now.} derived from Equation~\ref{eq:E}, suggests that modest increases of slack space, i.e., $(1-F)$, have a very substantial positive effect on cleaning efficiency. This impact declines as the fraction of slack space increases. As $1-F$ increases, it is increasingly probable that a page will have more than one out-of-date version. These versions are already empty, so another update of their page does not increase the emptiness of their containing segment. A fill factor in the range of $.70$ to $.9$ has high effectiveness, with $E$ in the range of about $.2$ to $.5$ and cost in the range of $10$ to $3.9$.

\section{Skewed Update Distributions}
\label{sec:skewed}
 
\subsection{Gedanken Analysis}

It is unusual for updates to be uniformly distributed over all the data. In this section, we show that update rate differences can be exploited to reduce cleaning costs. Our analysis for this is artificial, and the way we set things up is all but impossible to realize ``in real life''. But this analysis shows the potential if we are able to treat data with different update rates differently.

Assume that we have two different sets of pages, each with uniformly distributed updates, but with the number of pages and update frequencies different for each set. Our analysis goal is to show that if we manage each of these sets of pages completely separately in their own spaces, the write cost for the two sets can be made less than the write cost for the union of the two sets, with uniformly distributed updates. A similar analysis was done in~\cite{SA13}.  We hold constant the total page set size and slack space fraction. But we allocate the slack space where it is most useful in reducing write cost. We assume age based cleaning for each set and will use the results of the preceding analysis, as captured in Table~\ref{tab:clean} to determine $E$.

The total cost for managing the two sets of data separately is the weighted cost for the two sets. Let $D_i$ be the fraction of total data size in set $i$, and $U_i$ be the relative frequency of updates in set $i$. Then weighted average cost for updates is 
\[
Cost_{Total} = \sum_{i=1}^{2} Cost_i * U_i 
\]

We need to relate cost to $F$, rather than simply to $E$ because we want our result to be a function of how we divide the ``slack'' space between the two sets. For this we use $E = R * (1-F)$, where $R$ is simply the ratio we observe between $E$ and $(1-F)$. We have computed $R$ in Table~\ref{tab:clean}. Thus, our cost becomes
\[
Cost = \frac{2}{E} \approx \frac{2}{R*(1-F)}
\]
and total cost for our two separate sets of pages becomes
\[
Cost_{Total} = \sum_{i=1}^{2} {\frac{2}{R_i*(1-F_i)}* U_i}  
\]

\subsection{Dividing Slack Space}

\begin{figure}
\centering
\includegraphics[width=0.5\linewidth]{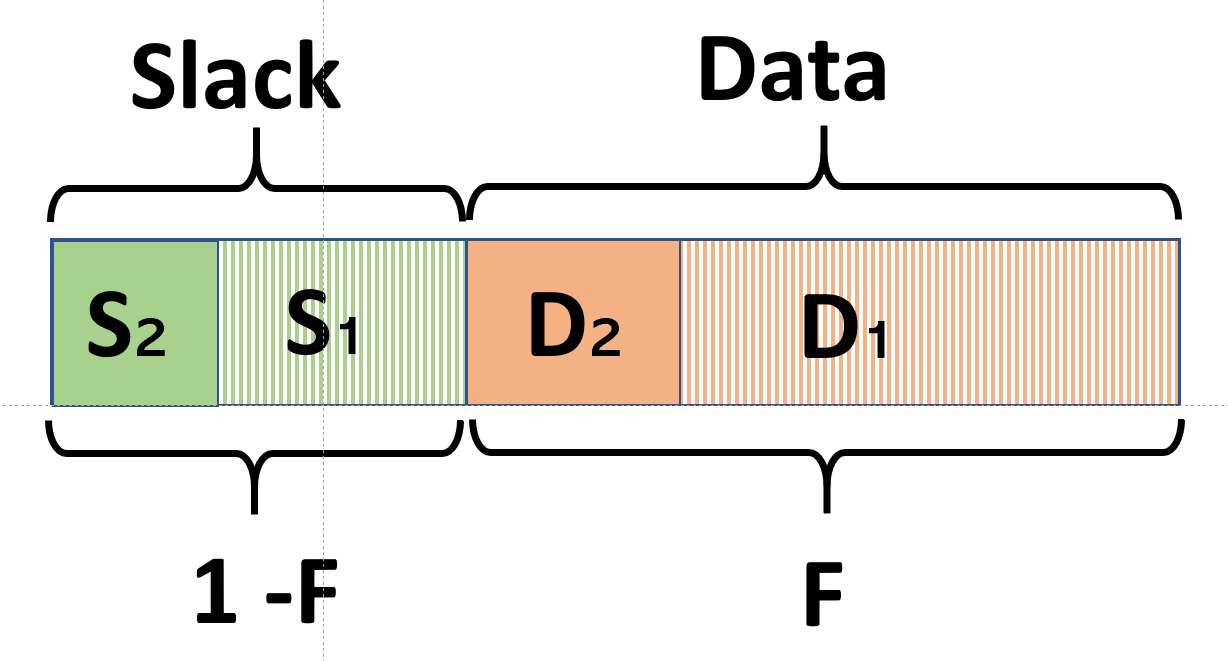}
\vspace{-0.1in}
\caption{How space is named and allocated when separating data by update frequency.}
\label{fig:data}
\end{figure}

We want to determine, for a given overall fill factor $F$, how to divide the slack space $(1 - F)$ so as to minimize the combined cleaning cost of our two data sets with differing update frequencies. We divide total space into two parts, each part devoted to managing either data set $1$ or data set $2$. $F = D_1 + D_2$ and slack space $(1 - F) = S_1 + S_2$. See Figure~\ref{fig:data}.

We are interested in dividing slack space so we set $S_i = (1-F)*g_i$ where $g_1 + g_2 = 1$.
Then we have
$$
F_i  = \frac{D_i}{D_i + S_i} = \frac{F*Dist_i}{(1-F)*g_i + F*Dist_i}
$$
where $Dist_i = D_i/(D_1 + D_2)$. We are normalizing our distribution and our computation here. For example, if we are dealing with an $80:20$ data distribution where $80\%$ of updates go to $20\%$ of the data, then $Dist_1 = 0.8$ and $Dist_2 = 0.2$. 

Plugging $F_i$ back into our cost equation yields
\[
Cost_i = \frac{2}{E_i} = \frac{2}{R_i} * ( 1 + \frac{F*Dist_i}{(1 - F)*g_i})*U_i
\]

We want to find the minimum for $Cost_{total}$, so we take the derivative of cost and set it to zero. We want cost in terms of a single variable $g_1$, so we replace $g_2$ by $1 - g_1$. We now make an important simplification- i.e. we assume that $R_i$ are constant. This is not true, but is a useful simplification as $R_i$ does not vary greatly. We hold update rate $U_i$, $F$, and $Dist_i$ constant. The total cost equation is then, abstractly, of the form 
\[
Cost_{Total} = A\frac{1}{g_1} + B\frac{1}{(1 - g_1)} 
\]
where all quantities in $A$ and $B$ as constants. This form permits us to easily see how to compute $d(Cost_{Total})/d(g_1)$.
\begin{small}
\[ 
\frac{d(Cost_{Total})}{d(g_1)} = -A \frac{1}{(g_1)^2} + B \frac{1}{(1-g_1)^2} = 0
\]
\end{small}

Replacing $A$ and $B$ by their update frequency and size and replacing $(1 - g_1)$ with $g_2$ results in 
\begin{small}
\[
(\frac{2*U_1}{R_1})*(\frac{F*Dist_1}{1-F})(\frac{2}{g_1^2}) = 
(\frac{2*U_2}{R_2})*(\frac{F*Dist_2}{1-F})(\frac{2}{g_2^2})
\]
\end{small}

We will solve the above equation for $g_1/g_2$. But we will simplify the equation for a special set of distributions so that we can clearly see a pattern in managing hot and cold data separately. The special set of distributions and update frequencies is the very frequently used $m:1-m$, where $m\%$ of the updates go to $1-m\%$ of the data. An example of this is the 80:20, where 80\% of the updates go to 20\% of the data. For these special distributions, 
$U_1 * Dist_1 = U_2 * Dist_2$. Canceling out these quantities as well as $F$ and $(1 - F)$ yield
\[
\frac{g_1} {g_2} = (\frac{R_2}{R_1})^{\frac{1}{2}}
\]
Importantly, slack space division is not related to the size of data being accommodated. 

Simplifying things even more, we note that $(R_2/R_1)^{1/2}) \approx 1$, so that, for these distributions, $g_1 \approx g_2$ and hence we simply share the slack space equally. But remember, this approximation applies only to our special set of distributions. 

\subsection{Minimum Cost}

We can now capture what would be the minimum cost for our two distributions, based on using the derived way above of sharing the slack space equally between hot and cold data. This is a very large, multi-parameter space. So we restrict ourselves to using a fill factor for the SSD of $.8$, and hence the slack space is $.2$. 
Dividing slack space equally has a differential impact on the fill factors for hot and cold data, with the hot data having a lower fill factor than the cold data. Said differently, we wait for hot segments to be emptier than cold segments. Thus, when update rates differ, we should not necessarily clean the emptiest segment first.

In Table~\ref{tab:min-cost}, we show the resulting minimum cost values for a number of distributions that are commonly used to evaluate performance. While we found the minimum cost to be achieved by splitting the slack space equally, the cost does not change very much over a range of space divisions. We include in Table~\ref{tab:min-cost} columns showing costs when the space is divided by giving hot data 60\% and cold data 40\%, and the reverse. This produces only very modest changes in cost, with cost values slightly higher than the minimum. 

\begin{table}
\begin{footnotesize}
\begin{center}
\begin{tabular}{|c|c|c|c|c|c|}
\hline
  F & Cold-Hot & MinCost & Hot:60\% & Hot:40\% & MDC-opt \\
\hline
0.8 & 90:10 & 2.96 & 3.06 & 2.99 & 2.96 \\
0.8 & 80:20 & 4.00 & 4.12 & 4.11 & 3.99 \\
0.8 & 70:30 & 4.80 & 4.90 & 4.86 & 4.76 \\
0.8 & 60:40 & 5.23 & 5.38 & 5.38 & 5.23 \\
0.8 & 50:50 & 5.38 & 5.46 & 5.46 & 5.38 \\
\hline

\hline
\end{tabular}
\end{center}
\end{footnotesize}
\caption{Minimum Cost When Managing Hot and Cold Data Separately}
\label{tab:min-cost} 
\end{table}

What we have shown here is that, if we can take into account the hot and cold data distribution, that we can clean more efficiently than if we have uniformly distributed updating. This leads us, in the next section, to how we might efficiently clean data to exploit this opportunity.

\section{Cleaning Order}
\label{sec:order}
 
\subsection{Minimum Declining Cost}

Section 2 demonstrated how well we can do assuming a uniform 
distribution of page writes and a circular buffer. But as demonstrated in Section 3, we can do better if there is a non-uniform distribution of page writes as we can then effectively give more of the slack space to the hot data, improving overall cleaning efficiency even as we reduce cleaning efficiency for cold data. To exploit this, we need to separate hot data and cold data by clustering data into segments based on its update rate.

We exploit a concept that we call declining cost to establish cleaning priority among segments with differing update rates. If cleaning cost per page of a segment can be expected to decline greatly over some time period if we are willing to wait, then we should wait. Instead, we should clean a segment from which we expect very little decline in per page cleaning cost. Indeed, we should clean the segment from which we expect the smallest decrease in cleaning cost, so long as we can reclaim some space.
 
The segment cleaning cost is related to the number of empty page frames in the segments. From Equation~\ref{eq:IOseg} we know that the cleaning cost per page frame declines as more page frames become empty due to update activity.
 
The general problem is to determine the optimal processing order for items whose processing costs decline with time. We assume that object $i$ has processing cost $c_{i}(t_{n})$ at time $t_{n}$. We further assume that the rate of change in cost of processing is constant, i.e., ${dc_{i}(t)}/{dt}$ is a constant. This is not true for cleaning segments, but over a modest interval, it is a useful approximation.
 
Thus, at $t_{0}$, $i$'s cost to clean is $c_{i}(t_{0})$. Then for any time $t_{i}$, the cost of cleaning object i is given by
\[     
c_{i}(t_i) \approx c_{i}(t_{0}) + \frac{dc_{i}(t_0)}{dt} (t_i - t_{0})
\]
The total cost of processing a collection of $k$ objects, in some particular order is
\[
 Cost_{Total} = \sum_{i=1}^{k} c_{i}(t_{i}) \approx
 \sum_{i=1}^{k} c_{i}(t_{0}) +
       \sum_{i=1}^{k} \frac{dc_{i}(t_0)}{dt} (t_{i}- t_{0})
\]

The derivative $\frac{dc(t)}{dt}$ will be negative for segment cleaning. It eases the discussion to talk about cost decline, which is captured by $-\frac{dc	(t)}{dt}$. Thus, we rewrite $Cost_{Total}$ as
\[
Cost_{Total} \approx
\sum_{i=1}^{k} c_{i}(t_{0}) -
       \sum_{i=1}^{k} -\frac{dc_{i}(t_0)}{dt} (t_{i}- t_{0})
\]

$Cost_{Total}$ is minimized when the second sum on the right hand side is maximized. We want to show that this is true when the largest cost declines are multiplied by the largest time intervals.
 
{\bf Maximality Lemma:} Given two sets of positive numbers X and Y, with $||X|| = ||Y||$, 
\[\sum_{i=1}^{k}x_{i} y_{i}\] is maximized
when $X = \{x_{i}\}$ and $Y = \{y_{i}\}$ are both ordered in the same way, i.e., $x_i \ge x_j$ iff $y_i \ge y_j$.
 
Proof: See the ``Maximality Lemma" in the appendix.
 
Hence, it is best to process the objects experiencing the largest rates of cost decline last, and conversely, to process first the objects with the smallest rates of decline. The intuition here is that it pays to wait for the large cost declines, and to process objects whose costs won't decline much more first.
 
\subsection{Segment Cleaning}
 
The preceding result will apply only approximately as the rate 
of decline in cleaning cost per page is not constant. Thus, this is a first order approximation. We need to differentiate the cost per segment function. As before, $E$ denotes the fraction of the segment page frames that are empty. Further, $U_{pf}$ denotes the frequency of updates to pages in the segment. Recall that the cost of writing a segment is:
\[
   Cost = \frac{2}{E} = 2 E^{-1}
\]
and hence that cost decline is
\[
   \frac{d(Cost)}{du} = (\frac{-2}{ E^2})( \frac{dE}{du})
\]

We measure time not in clock time but in update count, where the ``clock'' has one "tick" per update. We will refer to "time" subsequently, but it should be understood that our clock is an update counter. Rate of change with respect to updates is unaffected by variations in system load, outages, etc. In particular, update frequency differences would be reduced by a system outage were we to use clock time, since that would add outage time to age computations, which would reduce the ratio between update frequencies.

We need to determine the value for $dE/du$. To get the rate at which $E$ is changing in a segment, we need to multiply the number of non-empty pages by the segment update frequency.  Let $U_{pf}$ denote the update frequency per page. The update frequency per segment is then the number of current pages times $U_{pf}$. Then,
\[
 \frac{dE}{du} \; \alpha \; \ (1 - E) U_{pf} \Delta_E 
\]
where $\Delta_E$ is the change in E produced by one update. Hence
\[
   \frac{d(Cost)}{du} \; \alpha \; \frac{-2 (1-E)}{E^2} U_{pf} \Delta_E
\]
 
\subsection{Update Frequency}
\label{sub:Upd}
The big picture of our cleaning approach, like earlier efforts, is to separate pages into segments based on update frequency. To do this, we need a simple way to estimate update frequency.  Keeping extensive statistics on page updates is expensive. So, similar to~\cite{RO92}, we use some form of ``age'' as an approximate (and highly uncertain) estimate of a page's update frequency.  

We want to know what fraction of updates are updates for a given segment. This will determine update frequency as a function of overall update rate. We define $u_{now}$ as the current update count, $u_{p1}$ as the time of the ultimate (last) segment update,and $u_{p2}$ as the time of the penultimate (next to last) update.

We have found that using $u_{now} - u_{p1}$ as the update period (inverse of update frequency) is very inaccurate. Instead, we want to take a sample of two intervals instead of one as the the average interval between updates for a segment. Then, for our segment of interest, 
\[
U_{pf} = \frac{2}{u_{now} - u_{p2}} 
\]
That is, there are two updates to our segment over an update interval of $u_{now} - u_{p2}$.

It is possible to extend this to include a larger interval containing more updates, e.g.
\[
U_{pf} = \frac{n}{u_{now} - u_{pn}} 
\]
but $U_{pf}$ may change, and we want to track these changes, so we should not average over too many updates. Subsequently, we average $U_{pf}$ over two update intervals. Then our estimated cost decline function is
\[
- \frac{d(Cost)}{du} \; \alpha \; \frac{2 (1-E)}{E^2} * \frac{2}{u_{now} - u_{p2}}* \Delta_E
\]

\subsection{Updates and $\Delta_E$}

$\Delta_E$ is the change in emptiness $E$ produced by an update to a page in the segment. For a fixed size page, this change in $E$ is $1/P$ where $P$ is the number of pages in the segment. This yields a minimum cost decline function
\[
- \frac{d(Cost)}{du} \; \alpha \; \frac{2 (1-E)}{E^2} * \frac{2}{u_{now} - u_{p2}}* \frac{1}{P}
\]
Since $1/P$ is a constant, it does not impact the cleaning order of segments.
 
The situation is more complicated when pages can be of variable size. Assume that a segment is $B$ bytes long. What is the impact of an update to page P on emptiness. It is something like $Size_{P}/B$. But what do we use for $Size_{P}$? Since we do not know the next page to be updated, we cannot know its size. However, we can approximate the size by computing the average size of the remaining pages that have not yet been updated. This is a value for $\Delta_E$ averaged over the remaining valid pages of the segment. 

Let $C$ be the number of pages that have not yet been updated. Further let $A$ be the amount of available free space in our segment of size $B$. Then, the average size of the remaining un-updated pages is $(B - A)/C$. This yields a more general form for declining cost of 
\[
- \frac{d(Cost)}{du} \; \alpha \; \frac{2 (1-E)}{E^2} * \frac{2}{u_{now} - u_{p2}}* \frac{(B - A)/C}{B}
\]

\subsection{Uniformly Distributed Updates}

Subsection~\ref{sub:Upd} discussed how to estimate update frequency. However, if we know that updates are uniform, then
$U_{pf} = c$, i.e. constant. Taking this into account, with declining cost, we have for our minimum declining cost (MDC)
\[
\textrm{Priority[MDC]} \; \alpha \; \frac{2 (1-E)}{E^2}*\Delta_E
\]

We define Priority[greedy] as ordering segments for cleaning based on how much empty space they have. Essentially, ``greedy'' cleans the segment with the highest value of $E$ first, which is the lowest value of $1/E$ when defining high priority by low number as we have been doing. Greedy priority is what age based cleaning yields when applied to a uniform distribution. That is, cleaning the oldest first will have the same result as cleaning the emptiest first because, at least with high probability, the oldest segment is also the emptiest. 

For a uniform update distribution, Priority[MDC] orders the segments for cleaning in exactly the same order as Priority[greedy]. This is seen by noting that
\[
  \frac{1-E}{E^{2}} = \frac{1}{E} (\frac{1}{E}-1)
\]
The factor $(1/E)-1)$ is ordered in the same way as $(1/E)$. Both of these factors are positive. Thus their product is ordered as $(1/E)$. 
Hence, for a uniform distribution, Priority[MDC] $\alpha$ Priority[greedy].

\section{How to Clean}
\label{sec:how}

\subsection{Cost Information}

\subsubsection{Segment Related Information}

Given the analysis for declining cost in the prior section, we can identify information that we will need to realize cleaning based on an MDC strategy. Some of this information will be system determined and constant, like $B$ above. However, some of it changes over time and will be different for each segment. This is the information that we need to capture for each segment:

\begin{itemize}[leftmargin=.2in,itemsep=-3pt]
\item $A$: Available (free) storage in a segment.
\item $C$: Number of pages containing current page state. 
\item $u_{p2}$: The next to last time of update of pages in the segment.

\end{itemize}

\subsubsection{Global Information}

There is also global information that we need also to be aware of. It consists of:

\begin{itemize}[leftmargin=.2in,itemsep=-3pt]
\item $B$: the size in bytes of each segment.
\item $u_{now}$: the current time as measured by update count.
\end{itemize}

\subsubsection{Transformed Declining Cost Equation}

Using the quantities above and the definitions of $E = A/B$ and average page size, we can compute our declining cost function as:
\[
- \frac{d(Cost)}{du} \; \alpha \; \frac{2 (1-A/B)}{(A/B)^2} * \frac{2}{u_{now} - u_{p2}}* \frac{(B - A)/C}{B}
\]

We can factor out and then drop constant factors as they do not impact the cost decline ordering. Then simplifying this equation yields
\[
- \frac{d(Cost)}{du} \; \alpha \; (\frac{(B - A)}{A})^2 * \frac{1}{C*(u_{now} - u_{p2})}
\]

\subsection{Maintaining Segment Information}

\subsubsection{Easy Quantities}
Some information is easy to maintain, regardless of whether the segment was created by user updates or by garbage collection. 
\begin{itemize}[leftmargin=.2in,itemsep=-3pt]
\item Available space $A$ is set to an initial value when the segment is first filled. Every time a page of the segment is updated such that a current page in the segment is over-written, the size of the page is subtracted from $A$.
\item Number of pages $C$ containing current state is set to an initial value when the segment is first filled. Every time a page of the segment is updated such that a current page in the segment is over-written, the count $C$ is reduced by one.
\end{itemize}

\subsubsection{Update Frequency}

Maintaining update frequency is more challenging than other segment related information. Our goal when responding to events like garbage collection or processing a new user write is to carry forward the update history as accurately as possible. 

\noindent
\textbf{Garbage Collection Writes.}
Each page to be GC'd comes from a segment where we have maintained $u_{p2}$.  This is the value we use for the page. When we include the page in a new segment that contains re-written pages from other segments, the value for $u_{p2}$ for the new segment is the average $u_{p2}$ for all pages written to it. How crude this is depends in part on how we organize the pages to be separated by update frequency (see subsection \ref{sub:sep}).

\noindent
\textbf{New User Writes.} There are two cases to deal with: 
\begin{description}[leftmargin=.2in,itemsep=-3pt]
\item[Non-first Write:] The old state of each re-written page has a $u_{p2}$ that can be found from its containing segment. We start by assuming that the prior $u_{p1}$ is midway between the current $u_{now}$ and $u_{p2}$. With a new update, the prior $u_{p1}$ then becomes the new $u_{p2}$. Thus
\[
new(u_{p2}) = old(u_{p2}) + .5 * (u_{now} - old(u_{p2})) 
\]
\item[First Write:] $u_{p2}$ will be assigned a relatively arbitrary value since we have no update history to guide us. Since pages mostly contain cold data, assigning a $u_{p2}$ that makes the page ``coldish'' is usually appropriate. Thus we set its $u_{p2}$ to the oldest value of $u_{p2}$ in the batch of pages we are processing for new user writes. As additional updates are made (see above), this starting point for $u_{p2}$ will approach the update frequency the page is experiencing.
\end{description}

\subsection{Separating by Update Frequency}
\label{sub:sep}

As we have seen in Section~\ref{sec:skewed}, cleaning performance is improved when we can separate data by update frequency, i.e. separate hot from cold data. Thus, we try to group together pages from segments as much as possible by their values for $u_{p2}$. Since $u_{p2}$ is maintained in our segment summary information, or is assigned to each newly updated page, we have the information needed for doing this. Thus, when assigning pages to segments, we first sort the collection of pages by their value of $u_{p2}$. This is similar to what the Berkeley paper~\cite{RO92} proposed, though their technique used $u_{p1}$ instead of $u_{p2}$. Sorting permits us to cluster pages by $u_{p2}$ when assigning the pages to new segments, both for user and GC writes. Thus, it is a technique for separating pages into segments by update frequency.

Note that the sorted ordering of page moves to new segments may be slightly different from the minimum declining cost order. However, minimum declining cost order determines the set of candidate segments to clean. The sorting by $u_{p2}$ is within this candidate set. 

\section{Cleaning Evaluation}
\label{sec:eval}
In this section, we experimentally evaluate the proposed MDC cleaning algorithm and compare it against other algorithms.
We first describe the experiment setup followed by the detailed evaluation results.

\subsection{Experiment Setup}
\subsubsection{Simulation Setup}
As in prior work~\cite{RO92,SA13}, we built a simulator to evaluate the various cleaning algorithms.
The major difference between the simulator and an actual system is that the former only writes page IDs instead of page contents.
Since we only care about the cleaning cost in the evaluation, the actual page contents do not matter here.
For all experiments, the page size was set at 4KB and the segment size was set at 2MB (with 512 pages).
The size of the simulated system was set at 100GB\footnote{The actual size does not impact the write amplification.}.
Cleaning is triggered when the number of free segments falls below 32. Unless otherwise noted, each cleaning cycle cleans 64 in use segments. This amortizes the cost of running the cleaning algorithm over a batch of segments, and also enables more effective separation of pages by update frequency.

\subsubsection{Performance Metric}
We measured the comparative effectiveness of various cleaning algorithm, in terms of write amplification $Wamp$, as defined in equation~\ref{eq:Wamp}.
A good cleaning algorithm should have a low write amplification.
For example, if write amplification is 0, then all of the I/O bandwidth can be used for user writes.
In contrast, if the write amplification is 1, then only half of the I/O bandwidth can be used for user writes, while the other half is used for cleaning. This added toll of extra writes reduces the lifetime of the flash storage.

\subsubsection{Evaluated Algorithms}
We evaluated the following cleaning algorithms in our experiments.
The age-based algorithm (denoted as ``age") always selects the oldest segment for cleaning.
The greedy algorithm (denoted as ``greedy") always cleans the segment with the most available free space.
The cost-benefit algorithm~\cite{RO92} (denoted as ``cost-benefit") cleans segments with the largest cost benefits, which are defined as $\frac{(1-E)\times age}{E}$.

The multi-log algorithm~\cite{SA13}, which is the current state-of-the-art cleaning algorithm, separates pages into multiple logs so that pages within each log have similar update frequencies. We further evaluated two variations of the multi-log algorithm. The first variation (denoted as ``multi-log") estimates the page update frequency using the previous update timestamp. The second variation (denoted as ``multi-log-opt") uses the exact page update frequency to measure the best possible write amplification of the multi-log algorithm. For both variations, we only cleaned one segment at a time in order to be consistent with the evaluation conducted by~\cite{SA13}.

Similarly, we evaluated two variations of our proposed MDC algorithm. The first variation (denoted as ``MDC'') uses update timestamps to estimate page update frequencies and sorts pages to group pages with similar update frequencies together. The second variation (denoted as ``MDC-opt'') also uses the exact page update frequency for cleaning, as in multi-log-opt.

\subsubsection{Workloads}
We used two sets of workloads in the evaluation.
First, we used a synthetic workload where page writes were generated following a random distribution.
Second, we used the I/O traces collected from running the TPC-C benchmark on a \btree-based storage engine.
The detailed setup of these two workloads are further described below.

\subsection{Synthetic Workloads}
We first used synthetic workloads to evaluate various cleaning algorithms, where page writes were generated following predefined distributions as discussed below.
For each experiment, we wrote 10TB data, which is 100 times the size of the simulated log-structured system, so that the write amplification stabilized.
Throughout this evaluation, we focus on the following two questions:
First, what is the effectiveness of each optimization used in the proposed MDC algorithm?
Second, how well the proposed MDC algorithm performs compared to other cleaning algorithms?
In the remainder of this section, we answer these two questions through our evaluation.

\begin{figure}
	\centering
	\includegraphics[width=0.75\linewidth]{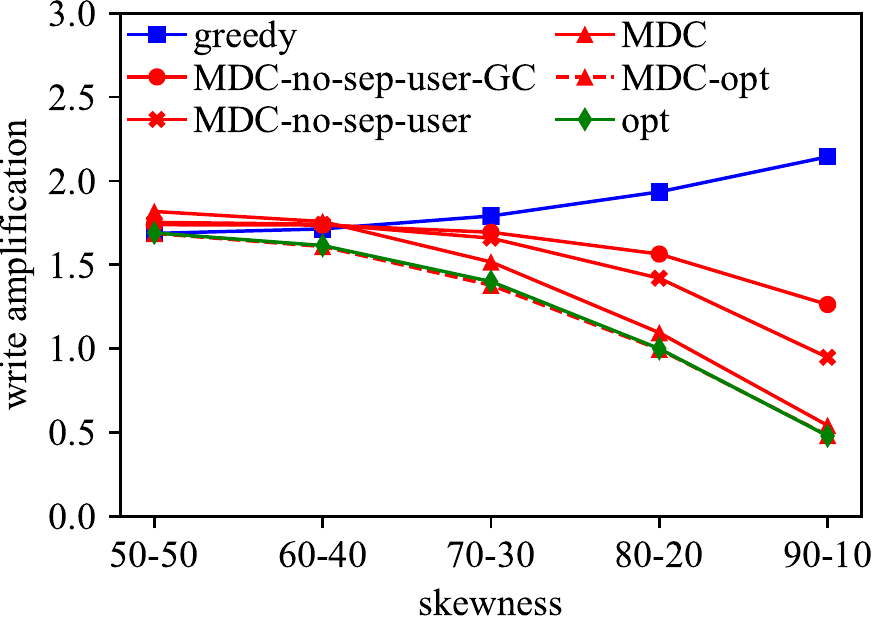}
	\vspace{-0.1in}
	\caption{Experimental Results on Hot-Cold Distributions}
	\label{fig:expr-breakdown}
\end{figure}

\begin{figure}
	\centering
	\includegraphics[width=0.75\linewidth]{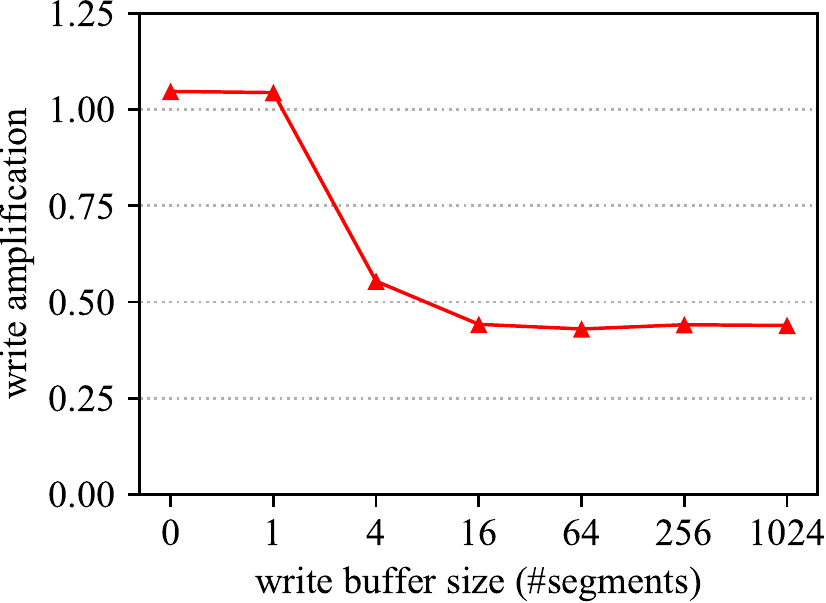}
	\vspace{-0.1in}
	\caption{Cleaning Impact of Sort Buffer Size}
	\label{fig:expr-sort}
\end{figure}

\begin{figure*}
	\centering
	\includegraphics[width=\linewidth]{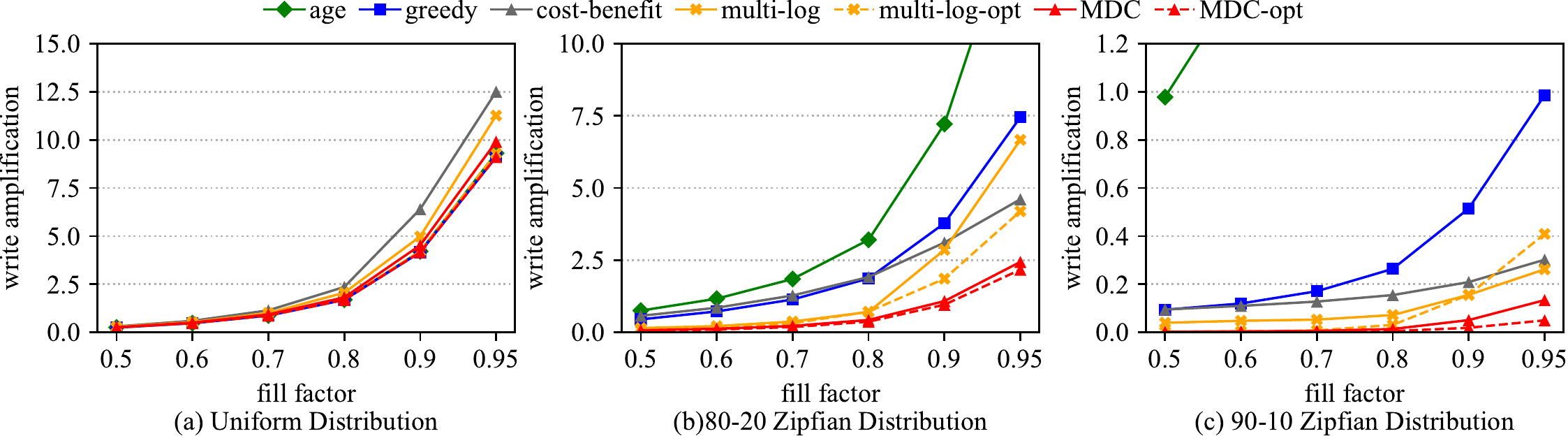}
	\vspace{-0.2in}
	\caption{Experimental Results on Synthetic Workload}
	\label{fig:expr-synthetic}
\end{figure*}

\subsubsection{Breakdown Analysis.}
To understand the effectiveness of each optimization used in the MDC algorithm, we carried out a breakdown analysis by removing these optimizations one by one as follows.
Here we used the same hot-cold distribution discussed in Section~\ref{sec:skewed}
where pages are divided into two groups with different sizes and update frequencies. The fill factor was set at 0.8.
The optimal write amplification (denoted as ``opt'') for this distribution has been shown in Table~\ref{tab:min-cost}.
In addition to MDC and MDC-opt, we further included the following variations by removing some optimizations from MDC:
\begin{itemize}[leftmargin=.2in,itemsep=-3pt]
\item MDC-no-sep-user: the variation of MDC that does not separate user writes by their update frequencies;
\item MDC-no-sep-user-GC: the variation of MDC that does not separate both user and GC writes by their update frequencies.
\end{itemize}
Finally, we further included the greedy algorithm in this evaluation.
Note that the only difference between greedy and MDC-no-sep-user-GC is that they use different criteria for selecting segments to clean.

The resulting write amplifications under different skewness are shown in \reffigure{fig:expr-breakdown}.
When the workload is not skewed (50-50), the greedy algorithm achieves the optimal write amplification, which is also the same as the opt algorithm.
In this case, the MDC and its variations incur some additional overhead due to estimation errors.
However, note that most practical workloads do not have uniform update distribution.
When the workload is skewed, the MDC and its variations start to outperform the greedy algorithm
because the greedy algorithm always cleans the segment with the most available space, which will leave cold segments uncleaned for a long time~\cite{RO92}.

The performance differences among the MDC and its variations also illustrate the effectiveness of each optimization.
By comparing MDC-no-seg-user and MDC-no-seg-user-GC under skewed workloads, e.g., 80-20, we can see that separating user writes is more effective that separating GC writes.
This is because most of the GC writes are cold pages but user writes contain a mix of cold and hot pages.
Moreover, the write amplification of MDC-opt also aligns with that of the opt algorithm as obtained via theoretical analysis.
This confirms the optimality of the proposed MDC algorithm.

We further carried out an experiment to evaluate the impact of the buffer size for sorting user writes
on the cleaning cost of the MDC algorithm.
Here we used the 80-20 Zipfian update distribution with the Zipfian factor 0.99.
Note that the Zipfian distribution is more complex and realistic than the hot-cold distribution because under the Zipfian distribution all pages have unique update frequencies.
The fill factor was again set at 0.8.
The resulting write amplifications under different buffer sizes (in terms of the number of segments) are shown in \reffigure{fig:expr-sort}.
As the figure shows, it is important to separate page writes based on their update frequencies,
as sorting page writes by their update frequencies significantly reduces the write amplification.
Moreover, using a write buffer with 16 segments (32MB in our evaluation) already achieves near-optimal write amplification.

\subsubsection{Comparison with Other Cleaning Algorithms.}
We used the uniform and Zipfian update distributions to evaluate the write amplification of various cleaning algorithms.
For the Zipfian distribution, we further evaluated two Zipfian factors representing different skewness,
namely the 80-20 Zipfian distribution (Zipfian factor 0.99) and the 90-10 Zipfian distribution (Zipfian factor 1.35).

The write amplification of the evaluated cleaning algorithms under different fill factors are shown in \reffigure{fig:expr-synthetic}.
Under the uniform distribution (\reffigure{fig:expr-synthetic}a), both the age-based and greedy algorithms have the optimal write amplification.
Moreover, both multi-log-opt and MDC-opt have achieved lowest write amplification as well.
In both algorithms, all pages are placed into one group because they all have the same update frequency.
It should be noted that the multi-log-opt algorithm always cleans the oldest segment within this group,
which behaves exactly as the age-based algorithm.
However, without the exact page update frequency, both multi-log and MDC incur some additional overhead due to estimation errors.
Moreover, the multi-log leads to a higher write amplification than MDC because it creates a large number of of logs during runtime, 
even though all pages have the same update frequency.
Finally, the cost-benefit algorithm has a much higher write amplification because it is purely a heuristic approach only optimized for skewed workloads.

Under the skewed Zipfian distributions (Figures \ref{fig:expr-synthetic}b and \ref{fig:expr-synthetic}c), the age-based algorithm produces a very high write amplification because it does not consider the update frequency of each page.
Moreover, the greedy algorithm also does not work well as discussed before.
The cost-benefit algorithm uses a heuristic to clean cold segments more aggressively by considering both the available space and the age, which reduces the write amplification.

By grouping pages with similar update frequencies, the multi-log algorithm further reduces the write amplification.
However, the multi-log algorithm still exhibits a high write amplification when the fill factor becomes larger.
Finally, the MDC algorithm achieves the lowest write amplification compared to all evaluated algorithms and has a similar write amplification as MDC-opt.

\begin{figure}
	\centering
	\includegraphics[width=0.7\linewidth]{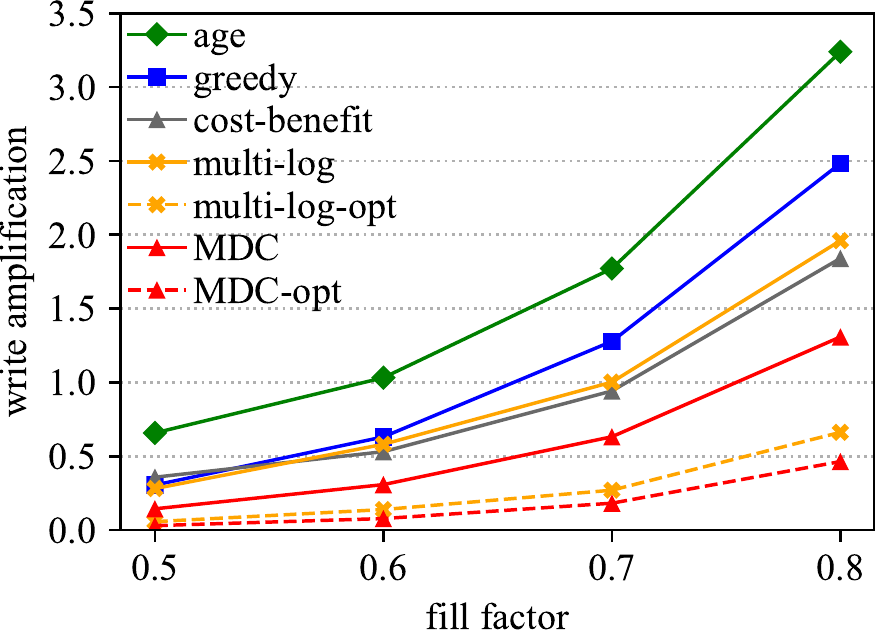}
	\vspace{-0.1in}
	\caption{Experimental Results on TPC-C Workload}
	\label{fig:expr-tpcc}
\end{figure}

\subsection{TPC-C Workloads}
To evaluate the cleaning algorithms on a somewhat more realistic workload, we carried out experiments using
the I/O traces collected from running the TPC-C benchmark on a \btree-based storage engine.
The buffer cache size was set at 4GB.
We varied the fill factor from 0.5 to 0.8, which corresponds to TPC-C scale factors 350 to 560.
It should be noted that the storage size under the TPC-C benchmark increases over time.
Thus, for each scale factor setting, we first loaded the TPC-C tables and ran the TPC-C benchmark until the fill factor increased by 0.1.
After collecting the I/O traces, we replayed them using the simulator to evaluate the cleaning algorithms.
The write amplification was measured during running phase, which contains about 250GB writes.

\reffigure{fig:expr-tpcc} shows the resulting write amplifications on the TPC-C traces.
In general, the TPC-C traces also contain hot pages and cold pages, and its skewness is close to 80-20~\cite{SA13}.
Thus, both the age-based and greedy algorithm do not work well, which is consistent with the results under the synthetic workloads.
By pre-analyzing page update frequencies, both multi-log-opt and MDC-opt achieve much lower write amplifications.
Moreover, MDC-opt further reduces the write amplification compared to multi-log-opt.

Without knowing page update frequencies in advance, both multi-log and MDC perform worse than their optimal alternatives.
Specifically, we saw that multi-log performs even slightly worse than the cost-benefit algorithm.
One reason is that multi-log initially places all pages into one log and adjusts the number of logs as the system runs.
Thus, its cleaning cost decreases over time but requires many writes before converging. However, the number of pages written in this experiment is much smaller than the previous experiment with synthetic workloads.

The TPC-C workload also has a shifting pattern where hot pages become cold over time~\cite{SA13}.
Since both algorithms use past update timestamps to estimate page update frequencies, this shifting pattern reduces the accuracy of the estimation.
Even so, MDC has a much lower write amplification compared to other cleaning algorithms under all fill factors.

\section{Related Work}

\subsection{Log Structured Stores.}
In addition to FTL in SSD controllers, log structured stores have been widely used in various data management components and systems.
Examples include key-value stores~\cite{rstore2020,llama},
NoSQL systems~\cite{oracle-nosql,aerospike2016}, in-memory stores~\cite{ramcloud2010,ramcloud2011},
log structured variants of B$^+$-trees~\cite{bwtree2013,btrfs2013},
and the key-value separation design~\cite{hashkv2018,hashkv2019,wicskey} for LSM-trees~\cite{lsm-survey,lsm1996}.
In these systems, cleaning is often the new bottleneck because of the expensive write amplification incurred by the cleaning process.
By minimizing the cleaning cost for log-structured stores, the proposed MDC algorithm will be able to further improve the performance of these systems.

\subsection{Cleaning Algorithms.}
The original LFS work~\cite{RO92} recognized the relationship between age and update rate. They also recognized that low update rate (old) segments should be cleaned at a lower value of $E$ than the more active update rate (young) segments. To capture this, they proposed a heuristic-based cleaning approach, called cost-benefit, that cleans cold segments more aggressively. However, as we have seen in Section~\ref{sec:eval}, the cost-benefit algorithm still fails to minimize the cleaning cost.
The LFS design and its cost-benefit algorithm has been adopted and refined by subsequent research~\cite{dac1999,marking2011,flashfs1995,WZ94}.
However, these cleaning algorithms are mainly based on heuristics, which fail to minimize the overall cleaning cost~\cite{SA13}.

Desnoyers~\cite{D12} proposed an analytical model to divide pages into a hot pool and a cold pool, which is similar to our discussion in Section~\ref{sec:skewed}.
In contrast, our MDC algorithm does not predefine the number of page groups. It instead sorts pages to group pages with similar update frequencies together
and selects the cleaning segments globally to minimize the overall cleaning cost.
Stoica and Ailamaki~\cite{SA13} proposed a multi-log approach to minimize the cleaning overhead, which is the state-of-the-art cleaning algorithm.
Its key idea is to separate pages into multiple logs so that pages within each log have similar update frequencies.
When writing to a log $L$ causes the system to be nearly full, the algorithm selects a local-optimal log to clean from $L$ and its two neighbors.
However, the multi-log approach has a few drawbacks, as shown in our evaluation.
First, it requires a lot of page writes to converge to its optimal state.
Moreover, it may not achieve the minimum cleaning cost, as its selects cleaning segments locally.
The proposed MDC algorithm addresses these two problems and further reduces the cleaning cost.

\section{Discussion}
\subsection{Analysis-Simulation Agreement}

Simulations can give you results that may be difficult or impossible to fully confirm. However, when analysis can be applied to the same conditions as were used for a simulation, we want to be able to confirm that the results produced from simulations agree with those produced by analysis. We have done this for the results from conditions that we were able to analyze. 

\begin{description}[leftmargin=.2in,itemsep=-3pt]
\item[Uniform Distribution:] From Table~\ref{tab:clean}, we can compare column {\bf E}, our analysis results (for age based cleaning, which is optimal for uniform distribution, and column {\bf MDC-opt}, the results produced by simulation using MDC. These columns agree to two significant digits. Figure~\ref{fig:expr-synthetic}a graphs the simulation results, and shows that MDC produces the lowest costs among the cleaning methods we compare.
\item[Hot:Cold Separation:] From Table~\ref{tab:min-cost} we can compare our analysis of minimum cost cleaning for distributions divided between hot and cold data in the {\bf MinCost} with our simulation under the same conditions using MDC, the {\bf MDC-opt} column. Again, the results are in agreement to at least two significant digits, a confirmation that MDC achieves the best possible results under the admittedly contrived conditions.
Figure~\ref{fig:expr-breakdown} shows that MDC achieves the optimal results and has lower cleaning costs than the other methods. 
\end{description}

The above agreement between analysis and simulation reinforce the confidence we can place in our overall results, including those under more general experimental conditions which we were unable to analyze. These confirmations strongly suggests that it will be very difficult to improve on MDC.  

\subsection{Conclusions}

Cleaning efficiency is critical to the efficiency of any log structured store, whether it be a software system such as a log structured file system~\cite{RO92} or a ``firmware'' system in an SSD controller's FTL. For an FTL, there is the additional incentive to reduce the write amplification so as to reduce flash storage wear. Our approach described here is intuitive, and produces results when we simulated its use that matched optimal results derived via our analysis. While it may be possible to improve the details of the approach, there is only limited room for subsequent improvements if update frequency changes only slowly.  However, knowledge of workload may make it possible to better predict update frequency changes, and knowing update frequency, as shown in our simulations, can often improve results further. 

\appendix
\section{Appendix}

{\bf Maximality Lemma:} Given  two sets of positive numbers X and Y, with $||X|| = ||Y||$, 
\[\sum_{i=1}^{k}x_{i} y_{i}\] is maximized
when $X = \{x_{i}\}$ and $Y = \{y_{i}\}$ are both ordered in the same way, i.e., $x_i \ge x_j$ iff $y_i \ge y_j$.

{\bf Proof:} By contradiction.

Assume that the lemma is not true. Then there exists at least two element pairs such that $x_i > x_j$ with $y_i < y_j$ for some i and j. Note that the sum of the products of all other x's and y's with these two entry pairs removed is unchanged regardless of how we order these two pairs. 

The contradiction maintains that the sum is maximal with the old order. Thus if we pair $x_i$ with $y_j$ and $x_j$ with $y_i$, we expect that:
\[
(x_i * y_i + x_j * y_j) - (x_i * y_j + x_j * y_i) \ge 0
\]
Clustering the y terms by factoring out the x terms gives:
\begin{align*}
	  x_i * (y_i - y_j) - x_j * (y_j - y_i) & \ge 0\\
\textrm{or, } x_i * (y_i - y_j) + x_j * (y_i - y_j) & \ge 0
\end{align*}
But $y_i < y_j$, so the term $(y_i - y_j) < 0$. Hence, the product of it times the positive $x$ values is also less than zero, a contradiction.


\begin{thebibliography}{10}

\bibitem{oracle-nosql}
{Oracle NoSQL Database Cloud Service}.
\newblock \url{https://www.oracle.com/database/nosql-cloud.html}.

\bibitem{wikiFTL}
{Wikipedia. Flash memory controller.}
\newblock
 \url{https://en.wikipedia.org/wiki/Flash_memory_controller#Flash_Translation_Layer_(FTL)_and_Mapping}.

\bibitem{wikiWA}
{Wikipedia. Write amplification.}
\newblock \url{https://en.wikipedia.org/wiki/Write_amplification}.

\bibitem{AM10}
R.~K. Agarwal and M.~Marrow.
\newblock A closed-form expression for write amplification in nand flash.
\newblock {\em IEEE Globecom Workshops}, pages 1846--1850, 2010.

\bibitem{hashkv2018}
H.~H.~W. Chan et~al.
\newblock {HashKV}: Enabling efficient updates in {KV} storage via hashing.
\newblock In {\em USENIX ATC}, pages 1007--1019, 2018.

\bibitem{dac1999}
M.-L. Chiang, P.~C.~H. Lee, and R.-C. Chang.
\newblock Using data clustering to improve cleaning performance for plash
 memory.
\newblock {\em Softw. Pract. Exper.}, 29(3):267–290, Mar. 1999.

\bibitem{DBB15}
N.~Dayan, L.~Bouganim, and P.~Bonnet.
\newblock Modelling and managing {SSD} write-amplification.
\newblock {\em CoRR}, abs/1504.00229, 2015.

\bibitem{D12}
P.~Desnoyers.
\newblock Analytic models of {SSD} write performance.
\newblock {\em TOS}, 10(2), Mar. 2014.

\bibitem{marking2011}
X.~{Hu}, R.~{Haas}, and E.~{Evangelos}.
\newblock Container marking: Combining data placement, garbage collection and
 wear leveling for flash.
\newblock In {\em MASCOTS}, pages 237--247, 2011.

\bibitem{flashfs1995}
A.~Kawaguchi, S.~Nishioka, and H.~Motoda.
\newblock A flash-memory based file system.
\newblock In {\em TCON}, pages 155--164, USA, 1995.

\bibitem{rstore2020}
L.~Lersch et~al.
\newblock Enabling low tail latency on multicore key-value stores.
\newblock {\em PVLDB}, 13(7):1091–1104, 2020.

\bibitem{llama}
J.~Levandoski, D.~Lomet, and S.~Sengupta.
\newblock {LLAMA}: A cache/storage subsystem for modern hardware.
\newblock {\em PVLDB}, 6(10):877–888, 2013.

\bibitem{bwtree2013}
J.~J. Levandoski, D.~B. Lomet, and S.~Sengupta.
\newblock The {Bw-Tree}: A {B-tree} for new hardware platforms.
\newblock In {\em ICDE}, pages 302--313. IEEE, 2013.

\bibitem{hashkv2019}
Y.~Li et~al.
\newblock Enabling efficient updates in {KV} storage via hashing: Design and
 performance evaluation.
\newblock {\em ACM TOS}, 15(3):1--29, 2019.

\bibitem{Lo94}
D.~Lomet. 
\newblock{Reclaiming Space in a Log Structured File System}
\newblock{Unpublished Digital Technical Report}, 1994.


\bibitem{wicskey}
L.~Lu et~al.
\newblock {WiscKey}: Separating keys from values in {SSD}-conscious storage.
\newblock In {\em USENIX FAST}, pages 133--148, 2016.

\bibitem{lsm-survey}
C.~Luo and M.~J. Carey.
\newblock {LSM}-based storage techniques: a survey.
\newblock {\em The VLDB Journal}, 29(1):393--418, 2020.

\bibitem{lsm1996}
P.~O'Neil et~al.
\newblock The log-structured merge-tree ({LSM}-tree).
\newblock {\em Acta Inf.}, 33(4):351--385, 1996.

\bibitem{ramcloud2010}
J.~Ousterhout et~al.
\newblock The case for {RAMClouds}: scalable high-performance storage entirely
 in {DRAM}.
\newblock {\em ACM SIGOPS Operating Systems Review}, 43(4):92--105, 2010.

\bibitem{ramcloud2011}
J.~Ousterhout et~al.
\newblock The case for {RAMCloud}.
\newblock {\em CACM}, 54(7):121--130, 2011.

\bibitem{PGK88}
D.~A. Patterson, G.~Gibson, and R.~H. Katz.
\newblock A case for redundant arrays of inexpensive disks ({RAID}).
\newblock In {\em SIGMOD}, page 109–116, 1988.

\bibitem{btrfs2013}
O.~Rodeh, J.~Bacik, and C.~Mason.
\newblock Btrfs: The linux {B-tree} filesystem.
\newblock {\em ACM TOS}, 9(3):1--32, 2013.

\bibitem{RO92}
M.~Rosenblum and J.~K. Ousterhout.
\newblock The design and implementation of a log-structured file system.
\newblock In {\em SOSP}, SOSP ’91, page 1–15, 1991.

\bibitem{RO92a}
M.~Rosenblum and J.~K. Ousterhout.
\newblock The design and implementation of a log-structured file system.
\newblock {\em TOCS}, 10(1):26–52, Feb. 1992.

\bibitem{aerospike2016}
V.~Srinivasan et~al.
\newblock Aerospike: Architecture of a real-time operational {DBMS}.
\newblock {\em PVLDB}, 9(13):1389–1400, 2016.

\bibitem{SA13}
R.~Stoica and A.~Ailamaki.
\newblock Improving flash write performance by using update frequency.
\newblock {\em PVLDB}, 6(9):733–744, July 2013.

\bibitem{WZ94}
M.~Wu and W.~Zwaenepoel.
\newblock Envy: A non-volatile, main memory storage system.
\newblock In {\em ASPLOS}, ASPLOS, page 86–97, 1994.

\end{thebibliography}
\end{document}